\documentclass[a4paper,twoside]{article}

\usepackage{fancyhdr}
\usepackage{multicol}
\usepackage{upgreek}
\usepackage{indentfirst}
\usepackage{booktabs}
\usepackage{array,tabularx}
\usepackage{keyval,graphicx}
\usepackage{textcomp}
\usepackage{amssymb,bm,mathrsfs,bbm,amscd}
\usepackage[tbtags]{amsmath}
\usepackage{lastpage}
\usepackage[numbers,sort&compress]{natbib}  

\newcommand{\newsmall}{\fontsize{9pt}{0.8\baselineskip}\selectfont}

\DeclareMathSizes{10}{10}{6}{5}

\font\scten=euex10 at 10pt
\newcommand{\vint}{\mathop{{\vcenter{\hbox{\scten\char90}}}}}

\font\ddt=tcrm1000 at 4pt
\newcommand{\dt}{\mathop{{\vcenter{\hbox{\ddt\char136}}}}}     

\newcommand{\dsize}{\displaystyle}
\newcommand{\dao}{{\rm d}}

\newcommand{\thanksmark}{\textsuperscript{\,\rm{*}}}

\renewcommand{\,}{\hspace{0.125em plus 0.025em minus 0.025em}}

\newcommand{\ruledown}{\hfill\noindent{\lower.38cm\hbox{\rule{0.2pt}{0.4cm}}\rule{8.35cm}
              {0.2pt}}\vspace*{-0.5cm}}

\let\asas=\cite
\renewcommand\cite[1]{\raisebox{0.3mm}{\textsuperscript{\asas{#1}}}}

\makeatletter

\renewcommand{\thefootnote}{\fnsymbol{footnote}}

\renewcommand{\thanks}[1]{\thanksmark
    \protected@xdef\@thanks{\@thanks
        \protect\footnotetext[0]{\hspace*{-6pt}$*$\;#1}}}

\newcounter{email}
\newcommand{\email}[1]{%
    \protected@xdef\@thanks{\@thanks%
        \protect\footnotetext[0]{\hspace*{-8pt}\arabic{email})\,{E-mail:\,}#1}}%
        \stepcounter{email}}%

\renewcommand\footnoterule{
  \kern 1\p@
  \hrule \@width37mm
  \kern 8\p@}

\renewcommand\@makefntext[1]{%
    \parindent 1em%
    \noindent
    \hb@xt@2em{\hss\@makefnmark}#1}

\renewcommand\maketitle{\par
  \begingroup
    \renewcommand\thefootnote{\@fnsymbol\c@footnote}%
    \def\@makefnmark{\rlap{\@textsuperscript{\normalfont\@thefnmark}}}%
    \long\def\@makefntext##1{\parindent 1em\noindent
            \hb@xt@2em{%
                \hss\@textsuperscript{\normalfont\@thefnmark}}##1}%
    \if@twocolumn
      \ifnum \col@number=\@ne
        \@maketitle
      \else
        \twocolumn[\@maketitle]%
      \fi
    \else
      \newpage
      \global\@topnum\z@   
      \@maketitle
    \fi
  \@thanks
  \endgroup
  \setcounter{footnote}{0}%
  \global\let\thanks\relax
  \global\let\maketitle\relax
  \global\let\@maketitle\relax
  \global\let\@thanks\@empty
  \global\let\@author\@empty
  \global\let\@date\@empty
  \global\let\@title\@empty
  \global\let\title\relax
  \global\let\author\relax
  \global\let\date\relax
  \global\let\and\relax}

\renewcommand\@maketitle{%
  \begin{center}%
  \let \footnote \thanks
   \vspace*{0.5em}
    {\LARGE\bf \@title \par}%
    {\normalsize%
      \lineskip .5em%
      \vskip 2em%
      \begin{tabular}[t]{c}%
        \@author%
      \end{tabular}}%
  \end{center}}%
\makeatother

\newcommand{\danwei}[1]{%
  \begin{center}%
    \vskip -1em%
    \begin{center}%
      {\footnotesize #1}%
    \end{center}%
  \end{center}%
}%

\renewenvironment{abstract}%
  {\small\vspace{0.5mm}%
   \list{}{\rightmargin 2em%
           \leftmargin 2em}%
    \item{}{\bf Abstract}\hspace*{0.5em}\relax}%
   {\endlist}

\newenvironment{keyword}%
  {\small\vspace{1mm}%
    \list{}{\rightmargin 2em%
           \leftmargin 2em}%
                \item{}{\bf Key~words}\hspace*{0.5em}\relax }%
       {\endlist%
        }%

\newenvironment{pacs}%
  {\small%
    \list{}{\rightmargin 2em%
           \leftmargin 2em}%
                \item{}{\bf PACS}\hspace*{0.5em}\relax }%
       {\endlist%
        \vskip 6mm}%

\makeatletter

\renewcommand \thesection {\bf\@arabic\c@section}

\renewcommand\section{\@startsection {section}{1}{\z@}%
                                    {5mm \@plus.2ex \@minus .2ex}%
                                   {5mm \@plus.2ex \@minus .2ex}%
                                   {\normalfont\large\bfseries}}
\renewcommand\subsection{\@startsection{subsection}{2}{\z@}%
                                     {1.5ex \@plus .2ex}%
                                     {1.5ex \@plus .2ex}%
                                     {\normalfont\bfseries}}

\renewcommand\subsubsection{\renewcommand \thesection {\@arabic\c@section}
                            \@startsection{subsubsection}{3}{\z@}%
                                     {0.5ex}%
                                     {0.5ex}%
                                     {\normalfont}}

\renewcommand{\@biblabel}[1]{#1}
\renewcommand\refname{{\normalsize\bf References}}
\renewenvironment{thebibliography}[1]
     {\noindent\refname%
      \@mkboth{\MakeUppercase\refname}{\MakeUppercase\refname}%
      \footnotesize
      \list{\@biblabel{\@arabic\c@enumiv}}%
           {\settowidth\labelwidth{\@biblabel{#1}}%
            \leftmargin\labelwidth
            \advance\leftmargin\labelsep
            \@openbib@code
            \usecounter{enumiv}%
            \let\p@enumiv\@empty
            \renewcommand\theenumiv{\@arabic\c@enumiv}}%
      \setlength{\itemsep}{0mm}
      \setlength{\labelsep}{0.8em}
      \setlength{\parsep}{0mm}
      \setlength{\parskip}{0mm}
      \setlength{\topsep}{0mm}
      \setlength{\partopsep}{0mm}
      \clubpenalty4000
      \@clubpenalty \clubpenalty
      \widowpenalty4000%
      \sfcode`\.\@m}
     {\def\@noitemerr
       {\@latex@warning{Empty `thebibliography' environment}}%
      \endlist}

\newenvironment{mylabc}
                {%
                 \newsmall
                 \let\\\@centercr
                 \list{}{\itemsep      \z@
                         \itemindent   -1em%
                         \listparindent0em
                         \leftmargin   3em
                         \rightmargin  2em}
                         \item\relax}
                {\endlist}

                {\footnotesize
                 \vspace{-1mm}
                 \let\\\@centercr
                 \list{}{\itemsep      \z@
                         \itemindent   0em%
                         \listparindent0em
                         \leftmargin   2mm
                         \rightmargin  0em}
                         \item\relax}

                {\endlist}
\makeatother

\makeatletter

\renewcommand\caption[1]{%
\sbox\@tempboxa{\newsmall #1}%
\ifdim \wd\@tempboxa >\hsize
\begin{mylabc}
\vspace{-2mm}
{\small #1}%
\vskip 1mm%
\end{mylabc}
\else
\global \@minipagefalse
\vspace*{-2mm}
\hb@xt@\hsize{\hfil\box\@tempboxa\hfil}%
\vskip 1mm%
\fi}
\makeatother

\begin{document}

\title{Bayesian credible interval construction for\\ Poisson statistics\thanks{Supported by
NSFC (19991483, 10491303)}}

\author{%
ZHU Yong-Sheng$^{1)}$\email{zhuys@ihep.ac.cn}%
}
\maketitle

\danwei{%
(Institute of High Energy Physics, CAS, Beijing 100049, China)\\
}

\begin{abstract}
The construction of the Bayesian credible (confidence)
interval for a Poisson observable including both the signal and background
with and without systematic uncertainties is presented. Introducing the
conditional probability satisfying the requirement of the background not
larger than the observed events to construct the Bayesian credible interval
is also discussed. A Fortran routine, BPOCI, has been developed to implement
the calculation.
\end{abstract}



\begin{multicols}{2}

\section{Introduction}
\vspace{-1mm}

The commonly accepted way to report errors on results in high energy physics
experiments is to present confidence intervals for the parameters to be
determined based on observed data. Very often, observables in an experiment
are Poisson variables including both
the signal and background processes, therefore, the construction of
confidence interval for parameter of signal in a Poisson process is of great
importance. For such a question, Conrad et al.\cite{1} reviewed the methods of
confidence belt construction in the frame of frequentist statistics, and
developed a FORTRAN program, POLE\cite{2}, to calculate the confidence intervals
for a maximum of observed events of 100 and a maximum signal expectation of
50. The ordering schemes for frequentist construction supported are the
Neyman method\cite{3}, likelihood ratio ordering\cite{4} and improved likelihood
ratio ordering\cite{5}. The systematic uncertainties in both the signal and
background efficiencies as well as systematic uncertainty of background
expectation have been taken into account in the confidence belt construction
by assuming a probability density function (pdf) which parameterizes our
knowledge on the uncertainties and integrating over this pdf. This method,
combining classical and Bayesian elements, is referred to as semi-Bayesian
approach.

In the frame of Bayesian statistics\cite{6}, Narsky\cite{7,8} depicted the
estimation of upper limits for Poisson statistic with the known background
expectation. Roe and Woodroofe proposed using a Bayes procedure with uniform
prior to determine credible intervals without inclusion of systematic
uncertainties of signal efficiency and background expectation\cite{9}. Treatment
of background uncertainty is discussed with the flat prior for simplified
cases of background expectation distributions in Refs.~[10,11]. Inclusion of
systematic uncertainties of signal efficiency and background expectation in
the upper limit calculation via Bayesian approach has been recently
discussed by ZHU\cite{12}.

In this paper, we describe the construction of Bayesian credible
(confidence) interval for a Poisson observable including both signal and
background with and without systematic uncertainties. Introducing the
conditional probability satisfying the requirement of the background not
larger than observed events to construct Bayesian credible interval is also
discussed. A Fortran program, BPOCI, has been developed to implement the
calculation\cite{13}.

\vspace{-1mm}
\section{Bayesian credible interval}
\vspace{-1mm}

 Throughout this paper we assume that in the signal window, where the signal
events (if exist) shall reside, the number of signal events is a Poisson
variable with unknown expectation $s$ to be inferred, and the number of
background events is a Poisson variable with known expectation $b$, the
conditional pdf of observing $n$ total events given $s$ is represented by a
Poisson probability
\begin{equation}
\label{eq1}
p(n\vert s)_b = {\rm e}^{-(s+b)}\frac{(s+b)^n}{n!}~.
\end{equation}

In the Bayesian statistics, any statistical inference on parameters is based
on posterior pdf. The posterior pdf in our question can be expressed as
\begin{equation}
\label{eq2}
h(s\vert n)=\frac{q(n\vert s)_b \pi (s)}{\vint\nolimits_0^{\,\infty} q(n\vert s)_b \pi (s)\dao s}~,
\end{equation}
where $q(n\vert s)_b $ is the conditional probability of observing $n$ total
events given $s$ signal events, and $\pi (s)$ is the non-informative prior
\begin{equation}
\label{eq3}
\pi(s)\propto \frac{1}{(s+b)^m}~,
\quad
s\geqslant 0, \quad b\geqslant 0 \quad 0\leqslant m\leqslant 1,
\end{equation}
where $m$=0 corresponds to Bayes prior, $m$=0.5 to $1 / \sqrt {s+b}$
prior, and $m$=1 to 1/($s+b)$ prior. The statistical bases on these three priors
are referred to Refs.~[14---17]. It should be kept in mind that different $m$
values will give different answers for the credible interval. The expected
coverage and length of Bayesian intervals constructed with these three
priors and of frequentist intervals with Neyman construction\cite{3} and unified
approach\cite{4} can be found in Ref.~[8]. It has been shown that the $1/\sqrt{s+b}$
prior is the most versatile
choice among the Bayesian methods, and it provides a reasonable mean
coverage for the credible interval for Poisson observable.

It is easy to build the central interval [$S_{\rm L}$, $S_{\rm U}$] for $s$ at a given
Bayesian posterior credible level $CL=1-\alpha$ by posterior pdf $h(s\vert n)$
\begin{equation}
\label{eq4}
\vint\nolimits_0^{\,S_{\rm L}} h(s\vert n)\dao s =
\frac{\alpha }{2} = \vint\nolimits_{S_{\rm U}}^{\,\infty} h(s\vert n) \dao s ~,
\end{equation}
and upper limit $S_{\rm UP}$
\begin{equation}
\label{eq5}
1-\alpha = \vint\nolimits_0^{S_{\rm UP}} {h(s\vert n)\dao s}~.
\end{equation}
However, the central interval and upper limit determined in such a way
suffers from the so-called ``flip-flopping'' policy, namely, to report a
central interval or an upper limit is artificially decided by the
experimenter's choice, which is similar to the Neyman construction\cite{3} and
certainly undesirable. To avoid this drawback, it is better to construct the
highest posterior density (HPD) credible interval $R$ expressed as
\begin{equation}
\label{eq6}
1-\alpha = \vint\nolimits_R h(s\vert n) \dao s~ ,
\end{equation}
where for any $s_1 \in R$ and $s_2 \notin R$, the following inequality holds
\begin{equation}
\label{eq7}
h(s_1 \vert n) \geqslant h(s_2 \vert n)~.
\end{equation}
The HPD interval is the optimum and shortest interval at a given credible
level in the Bayes framework. It automatically provides a two-sided interval
or an upper limit, decided by the observed data itself, which is the same as
in the unified approach\cite{4}, superior over the Neyman construction.

\subsection{Bayesian interval without inclusion of systematic uncertainties}

In the case that the systematic uncertainties of the signal efficiency and
background expectation can be neglected, the signal expectation $s$ is an
unknown constant and the background expectation $b$ is a known value. In this
case, $q(n\vert s)_b $ in Eq.~(\ref{eq2}) is simply equal to $p(n\vert s)_{b}$ in
Eq.~(\ref{eq1}), and the posterior pdf is then given by
\begin{equation}
\label{eq8}
h(s\vert n) = \frac{(s+b)^{n-m} {\rm e}^{-(s+b)}}{\varGamma (n-m+1,b)}~,
\end{equation}
where
\begin{equation}
\label{eq9}
\varGamma (x,b)=\vint\nolimits_b^{\,\infty} s^{x-1} {\rm e}^{-s} \dao s~, \quad x>0,~ b>0
\end{equation}
is an incomplete gamma function. Solving the eqation of
\[
\frac{\dao h(s\vert n)}{\dao s}=0
\]
gives the unique solution of $s_m =n-(b+m)$, where $s_m $is the maximum of
$h(s\vert n)$, which can be known from the behaviors of $p(n\vert s)_b $ and
$\pi (s)$. While the $h(s\vert n)$ is a monotonic decreasing function of $s$ for
$n\leqslant b+m$, it is a function of $s$ with a single maximum for $n>b+m$. The
bounds of the HPD interval $R$, [$S_{\rm L}$, $S_{\rm U}$], at a given credible level
$CL=1-\alpha$ can be acquired by solving Eq.~(\ref{eq6}) numerically with posterior
pdf Eq.~(\ref{eq8}) from the measured values of $n$ and $b$.

\subsection{Bayesian interval with inclusion of systematic uncertainties}

Now we take into account the systematic uncertainties. In this case, both
the signal expectation and background expectation are not the constants, but
the variables; they have respective distributions.

If only the uncertainty of background expectation is present, and the
distribution of the background expectation is represented by a pdf $f_{b'}(b,\sigma _b )$
with the mean $b$ and standard deviation $\sigma _b $, the
conditional pdf expressed by Eq.~(\ref{eq1}) now is modified to
\begin{equation}
\label{eq10}
q(n\vert s)_b = \vint\nolimits_0^{\,\infty} {p(n\vert s)_{b'} \dt f_{b'} (b,\sigma _b) \dao b'}~,
\end{equation}
where $p(n\vert s)_{b'} $has the same expression in Eq.~(\ref{eq1}) with $b$ replaced by ${b}'$.

Next we consider both the uncertainties of the signal efficiency and
background expectation exist, and assume they are independent of each other.
The distribution of the relative signal efficiency $\varepsilon $ (with
respect to the mean of the signal detection efficiency $\eta$) is expressed by a
pdf $f_\varepsilon (1,\sigma _\varepsilon )$ with the mean 1 and standard
deviation $\sigma _\varepsilon $. The conditional pdf described by Eq.~(\ref{eq1}) is
then further modified to
\begin{equation}
\label{eq11}
q(n\vert s)_b = \vint\nolimits_0^\infty \vint\nolimits_0^\infty  p(n\vert s\varepsilon)_{b'}
f_{b'} (b,\sigma _b) f_\varepsilon (1,\sigma _\varepsilon) \dao b' \dao \varepsilon~,
\end{equation}
where $p(n\vert s\varepsilon )_{{b}'} $ represents that $b$ is replaced by
${b}'$, and $s$ by $s\varepsilon $ in Eq.~(\ref{eq1}). One notices that the lower limits
of integrals in Eqs.~(10), (11) are all zero, which are the possible minimum
value of any efficiencies and number of background events. One can determine
the corresponding posterior pdf $h(s\vert n)$ according to Eq.~(\ref{eq2})\cite{18}, and
then the Bayesian interval on $s$ at any given credible level with inclusion of
systematic uncertainties in terms of Eq.~(\ref{eq6}).

\subsection{Connection to the ``Conditioning''}

In the unified approach, there is a background dependence of the confidence
interval for a Poison observable in the case of fewer events observed than
expected background. Roe and Woodroofe\cite{5} propose a solution to this
problem by using such an argument that, given an observation $n$, the background
$b$ can not be larger than $n$ in any case. Replacing the usual Poisson
probability by the conditional probability satisfying this requirement in
confidence interval construction is called ``Conditioning''. The conditional
probability of observing $n$ total events with background $b$ not larger then $n$ is
\begin{equation}
g(n\vert s)_b  =  \frac{p(n\vert s)_b}{\dsize\sum\nolimits_{i=0}^n p(i)_b}~,
\quad
p(i)_b = \frac{1}{i!} b^i {\rm e}^{-b}~.
\end{equation}
As pointed out by authors of Ref.~[5]
that in the case the systematic uncertainties of the signal efficiency and
background expectation can be neglected, $g(n\vert s)_b $ is a density in
$s$, and, is the posterior distribution that is obtained when $s$ is given an
(improper) uniform prior distribution over the range $0\leqslant s<\infty $. Using
$g(n\vert s)_b $ as the posterior function $h(s\vert n)$ in Eq.~(\ref{eq6}) leads to
a Bayesian credible interval with flat prior ($m$=0). Therefore, if the
systematic uncertainties are taken into account in the Bayesian approach,
$q(n\vert s)_b $ with conditioning can be expressed by Eqs.~(10), (11) with
$p(n\vert s)_b $ replaced by $g(n\vert s)_b $. Thus, the Bayesian interval
with conditioning can be built by solving Eq.~(\ref{eq6}) with this $q(n\vert s)_b $.
Notice the factors irrelevant to $s$ in the numerator and denominator of
$q(s\vert n)$ are cancelled out, the consequences of introducing the
conditioning are quite amusing: only the intervals with inclusion of the
systematic uncertainty of background expectation (and also the systematic
uncertainty of signal efficiency simultaneously) differ from those without
conditioning, while the intervals kept unchanged in other cases. Besides,
introducing the conditioning produces shorter Bayesian intervals.

\section{BPOCI: An algorithm for calculating Bayesian interval}

We have developed an algorithm for calculating Bayesian interval for the
Poisson observable at a given credible level with or without inclusion of
syste-matic uncertainties in background (bkgd) expectation and signal
efficiency. It has been implemented as a\linebreak
FORTRAN program, BPOCI (Bayesian
POissonian Credible Interval)\cite{13}.

To run BPOCI, following valuables are required to input:\\[-2mm]
\[
\rm IC,\, II,\, ID,\, IBK,\, IE,\, N,\, B,\, SIGBK,\, SIGE,\, ETA,\, CL,\, AM.
\]

\noindent
Their meanings are listed in Table~1.

\vspace{2mm}

\begin{center}
\caption{Table~1.\quad Input variables of BPOCI and their meanings.}
\footnotesize
\begin{tabular*}{80mm}{@{\extracolsep{\fill}}cl}
\toprule
notation & \hspace*{7em}meaning \\
\hline
IC   & flag for conditioning: 1---no conditioning,\\
     & 2---conditioning. \\
II   & flag for type of interval: 1---HPD, 2---central,\\
     & 3---upper limit.  \\
ID   & flag for systematic uncertainties. (see below) \\
IBK  & flag for selecting the distribution of bkgd \\
     & expectation. (see below)  \\
IE   & flag~ for~ selecting~ the~ distribution~ of~ signal \\
     & detection efficiency. (see below) \\
N    & number~ of~ total~ events~ observed~ in~ signal \\
     & window, $n$. \\
B    & predicted bkgd expectation in signal window, $b$. \\
SIGBK& standard deviation of the distribution for relative\\
     & bkgd expectation, $\sigma_b/b$. \\
SIGE & standard deviation of the distribution for relative \\
     & signal efficiency, $\sigma _\varepsilon $. \\
ETA  & predicted signal detection efficiency, $\eta $. \\
CL   & credible level, $CL=1-\alpha $. \\
AM   & prior selection. $AM=m$, prior is 1/$(s+b)^{m}$,\\
     & 0$\leqslant m\leqslant $1. \\
\bottomrule
\end{tabular*}%
\vspace{2mm}
\end{center}

Flag ID can take four values with the following assignment:

1~ ---~ without considering any systematic uncertainties,

2~ ---~ incorporating~ systematic~ uncertainty~ of bkgd expectation,

3~ ---~ incorporating~ systematic~ uncertainty~ of\linebreak   signal efficiency,

4~ ---~ incorporating systematic uncertainties of bkgd expectation and signal
efficiency simultaneously, and they are assumed to be independent of each
other.

For the distribution of relative signal efficiency (signal efficiency
divided by $\eta$) or relative background expectation (background expectation
divided by $b$), three types of functions with the mean 1 and standard
deviation $\sigma $ are supported: Gaussian, Log-Gaussian and flat
distributions, which correspond to IE (IBK) equal to 1, 2 and 3,
respectively.

The program will automatically generate an output file ``BPOCI.out'', which
gives the bounds of the required interval at a given credible level.
Simultaneously, it creates a HBOOK file ``bpoci.hbk'', which can be looked
at by using PAW and to generate a corresponding ``bpoci.eps'' file, drawing
the posterior density $h(s\vert n)$ as a function of $s$ with both linear and
log scales. Fig.~1 shows the posterior density for $n$=8, $b$=2 with Gaussian
systematic uncertainty ${\sigma _b}/b$=0.3 using uniform prior, the HPD
interval at $CL$=0.9 is [2.07, 11.77]. Tables 2---4 list Bayesian HPD intervals
for signal expectation $s$ without inclusion of systematic uncertainties using
uniform prior at the most commonly used 68.27{\%}, 90{\%} and 95{\%} $CL$.

The inclusion of systematic uncertainties leads to widening the credible
interval. The relations of the interval's length versus the type of priors,
the type of pdfs for uncertainties and the uncertainty's size, have similar
tendencies as those of Bayesian upper limit, which are referred to Ref.~[12].
\end{multicols}

\vspace{4mm}

\begin{center}
\caption{Table~2.\quad Bayesian credible intervals for signal expectation $s$ without
incorporating systematic uncertainties using flat prior. $n$ and $b$ represents
the total events observed and background events, respectively.}
\footnotesize
\begin{tabular*}{170mm}{@{\extracolsep{-2.2mm}}ccccccccccc}
\multicolumn{11}{c}{68.27\% $CL$}\\
\toprule
$n\backslash b$ & 0.0 & 0.5 & 1.0 & 1.5 & 2.0 & 2.5 & 3.0 & 3.5 & 4.0 & 5.0 \\
\hline
 0 & 0.00,1.15 & 0.00,1.15 & 0.00,1.15 & 0.00,1.15 & 0.00,1.15
   & 0.00,1.15 & 0.00,1.15 & 0.00,1.15 & 0.00,1.15 & 0.00,1.15 \\
 1 & 0.27,2.50 & 0.00,1.99 & 0.00,1.79 & 0.00,1.66 & 0.00,1.57
   & 0.00,1.51 & 0.00,1.46 & 0.00,1.42 & 0.00,1.39 & 0.00,1.35 \\
 2 & 0.87,3.85 & 0.38,3.30 & 0.00,2.66 & 0.00,2.37 & 0.00,2.16
   & 0.00,2.01 & 0.00,1.89 & 0.00,1.80 & 0.00,1.72 & 0.00,1.61 \\
 3 & 1.56,5.14 & 1.06,4.64 & 0.58,4.08 & 0.16,3.42 & 0.00,2.95
   & 0.00,2.68 & 0.00,2.47 & 0.00,2.30 & 0.00,2.16 & 0.00,1.96 \\
 4 & 2.29,6.40 & 1.79,5.90 & 1.30,5.39 & 0.83,4.83 & 0.39,4.20
   & 0.00,3.52 & 0.00,3.20 & 0.00,2.94 & 0.00,2.73 & 0.00,2.40 \\
 5 & 3.06,7.63 & 2.56,7.13 & 2.06,6.63 & 1.57,6.11 & 1.10,5.56
   & 0.65,4.96 & 0.24,4.30 & 0.00,3.73 & 0.00,3.43 & 0.00,2.96 \\
 6 & 3.85,8.83 & 3.35,8.34 & 2.85,7.83 & 2.35,7.33 & 1.86,6.82
   & 1.38,6.27 & 0.93,5.69 & 0.51,5.06 & 0.12,4.38 & 0.00,3.64\\
 7 & 4.65,10.03& 4.15,9.53 & 3.65,9.03 & 3.15,8.53 & 2.66,8.02
   & 2.16,7.51 & 1.69,6.97 & 1.23,6.40 & 0.80,5.79 & 0.01,4.45 \\
 8 & 5.47,11.21& 4.97,10.71& 4.47,10.21& 3.97,9.71 & 3.47,9.21
   & 2.98,8.70 & 2.48,8.19 & 2.00,7.66 & 1.54,7.10 & 0.68,5.88 \\
 9 & 6.30,12.37& 5.80,11.88& 5.30,11.38& 4.80,10.88& 4.30,10.38
   & 3.80,9.87 & 3.30,9.37 & 2.81,8.86 & 2.33,8.33 & 1.41,7.21 \\
10 & 7.14,13.54& 6.64,13.04& 6.14,12.54& 5.64,12.04& 5.14,11.54
   & 4.64,11.04& 4.14,10.53& 3.64,10.03& 3.15,9.52 & 2.19,8.46 \\
11 & 7.99,14.69& 7.49,14.19& 6.99,13.69& 6.49,13.19& 5.99,12.69
   & 5.49,12.19& 4.99,11.69& 4.49,11.19& 3.99,10.69& 3.01,9.66 \\
12 & 8.84,15.83& 8.34,15.33& 7.84,14.83& 7.34,14.33& 6.84,13.83
   & 6.34,13.33& 5.84,12.84& 5.34,12.33& 4.84,11.83& 3.85,10.82\\
13 & 9.70,16.98& 9.20,16.47& 8.70,15.97& 8.20,15.47& 7.70,14.97
   & 7.20,14.47& 6.70,13.97& 6.20,13.47& 5.70,12.97& 4.70,11.97\\
14 &10.56,18.11&10.06,17.61& 9.56,17.11& 9.07,16.61& 8.57,16.11
   & 8.06,15.61& 7.57,15.11& 7.07,14.61& 6.56,14.11& 5.57,13.11\\
15 &11.43,19.24&10.94,18.74&10.43,18.24& 9.93,17.74& 9.44,17.24
   & 8.94,16.74& 8.43,16.24& 7.93,15.74& 7.43,15.24& 6.43,14.24 \\
16 &12.31,20.37&11.81,19.86&11.31,19.36&10.81,18.87&10.31,18.36
   & 9.81,17.87& 9.31,17.37& 8.81,16.86& 8.31,16.37& 7.31,15.37  \\
17 &13.18,21.49&12.68,20.99&12.18,20.49&11.69,19.99&11.19,19.49
   &10.68,18.98&10.18,18.49& 9.69,17.99& 9.19,17.49& 8.19,16.49 \\
18 &14.07,22.61&13.57,22.10&13.07,21.60&12.57,21.11&12.07,20.61
   &11.57,20.11&11.07,19.60&10.57,19.11&10.07,18.61& 9.07,17.61  \\
19 &14.95,23.72&14.45,23.22&13.95,22.72&13.45,22.22&12.95,21.72
   &12.45,21.22&11.95,20.72&11.45,20.22&10.95,19.72& 9.95,18.72  \\
20 &15.84,24.83&15.34,24.33&14.84,23.83&14.34,23.33&13.84,22.83
   &13.34,22.34&12.84,21.83&12.34,21.33&11.84,20.83&10.84,19.84 \\
\hline
\hline
$n\backslash b$ & 6.0 & 7.0 & 8.0 & 9.0 &10.0 &11.0 &12.0 &13.0 &14.0 &15.0 \\
\hline
 0 & 0.00,1.15 & 0.00,1.15 & 0.00,1.15 & 0.00,1.15 & 0.00,1.15
   & 0.00,1.15 & 0.00,1.15 & 0.00,1.15 & 0.00,1.15 & 0.00,1.15 \\
 1 & 0.00,1.32 & 0.00,1.30 & 0.00,1.28 & 0.00,1.27 & 0.00,1.26
   & 0.00,1.25 & 0.00,1.24 & 0.00,1.23 & 0.00,1.23 & 0.00,1.22 \\
 2 & 0.00,1.54 & 0.00,1.48 & 0.00,1.44 & 0.00,1.41 & 0.00,1.38
   & 0.00,1.36 & 0.00,1.34 & 0.00,1.33 & 0.00,1.32 & 0.00,1.30 \\
 3 & 0.00,1.82 & 0.00,1.72 & 0.00,1.64 & 0.00,1.58 & 0.00,1.53
   & 0.00,1.50 & 0.00,1.47 & 0.00,1.44 & 0.00,1.42 & 0.00,1.40 \\
 4 & 0.00,2.17 & 0.00,2.00 & 0.00,1.88 & 0.00,1.79 & 0.00,1.71
   & 0.00,1.65 & 0.00,1.61 & 0.00,1.57 & 0.00,1.53 & 0.00,1.50 \\
 5 & 0.00,2.61 & 0.00,2.36 & 0.00,2.18 & 0.00,2.04 & 0.00,1.93
   & 0.00,1.84 & 0.00,1.77 & 0.00,1.71 & 0.00,1.67 & 0.00,1.62 \\
 6 & 0.00,3.16 & 0.00,2.81 & 0.00,2.54 & 0.00,2.34 & 0.00,2.19
   & 0.00,2.06 & 0.00,1.97 & 0.00,1.89 & 0.00,1.82 & 0.00,1.76 \\
 7 & 0.00,3.83 & 0.00,3.35 & 0.00,2.99 & 0.00,2.71 & 0.00,2.50
   & 0.00,2.33 & 0.00,2.19 & 0.00,2.09 & 0.00,2.00 & 0.00,1.92 \\
 8 & 0.00,4.61 & 0.00,4.01 & 0.00,3.53 & 0.00,3.16 & 0.00,2.88
   & 0.00,2.65 & 0.00,2.47 & 0.00,2.32 & 0.00,2.20 & 0.00,2.10 \\
 9 & 0.57,5.96 & 0.00,4.77 & 0.00,4.18 & 0.00,3.70 & 0.00,3.33
   & 0.00,3.03 & 0.00,2.79 & 0.00,2.60 & 0.00,2.45 & 0.00,2.32 \\
10 & 1.29,7.31 & 0.48,6.04 & 0.00,4.92 & 0.00,4.34 & 0.00,3.86
   & 0.00,3.48 & 0.00,3.18 & 0.00,2.93 & 0.00,2.73 & 0.00,2.56 \\
11 & 2.06,8.58 & 1.19,7.40 & 0.39,6.10 & 0.00,5.06 & 0.00,4.49
   & 0.00,4.01 & 0.00,3.63 & 0.00,3.32 & 0.00,3.06 & 0.00,2.85 \\
12 & 2.88,9.78 & 1.95,8.68 & 1.09,7.48 & 0.31,6.16 & 0.00,5.20
   & 0.00,4.63 & 0.00,4.16 & 0.00,3.77 & 0.00,3.45 & 0.00,3.19 \\
13 & 3.71,10.95& 2.75,9.90 & 1.84,8.77 & 1.00,7.55 & 0.23,6.22
   & 0.00,5.34 & 0.00,4.77 & 0.00,4.30 & 0.00,3.91 & 0.00,3.58 \\
14 & 4.57,12.10& 3.59,11.07& 2.64,10.00& 1.74,8.86 & 0.92,7.62
   & 0.16,6.28 & 0.00,5.47 & 0.00,4.91 & 0.00,4.44 & 0.00,4.04 \\
15 & 5.44,13.24& 4.44,12.22& 3.47,11.19& 2.53,10.10& 1.65,8.94
   & 0.84,7.68 & 0.10,6.34 & 0.00,5.59 & 0.00,5.04 & 0.00,4.57 \\
16 & 6.31,14.36& 5.31,13.36& 4.33,12.34& 3.36,11.29& 2.43,10.19
   & 1.56,9.01 & 0.76,7.74 & 0.03,6.39 & 0.00,5.72 & 0.00,5.17  \\
17 & 7.18,15.49& 6.18,14.48& 5.20,13.48& 4.21,12.45& 3.26,11.39
   & 2.34,10.27& 1.48,9.08 & 0.70,7.80 & 0.00,6.47 & 0.00,5.84 \\
18 & 8.07,16.61& 7.07,15.61& 6.07,14.60& 5.08,13.59& 4.10,12.56
   & 3.16,11.48& 2.25,10.35& 1.41,9.14 & 0.63,7.85 & 0.00,6.58 \\
19 & 8.95,17.72& 7.95,16.72& 6.95,15.72& 5.96,14.71& 4.97,13.69
   & 4.00,12.65& 3.06,11.57& 2.17,10.43& 1.34,9.20 & 0.57,7.91 \\
20 & 9.84,18.84& 8.84,17.83& 7.84,16.83& 6.84,15.83& 5.85,14.82
   & 4.86,13.80& 3.90,12.75& 2.97,11.66& 2.09,10.50& 1.27,9.26 \\
\bottomrule
\end{tabular*}%
\end{center}

\clearpage

\vspace{3mm}

\begin{center}
\caption{Table~3.\quad 90\% $CL$.}
\vspace{-3mm}
\footnotesize
\begin{tabular*}{170mm}{c@{\extracolsep{\fill}}ccccccc}
\toprule
$n\backslash b$ & 0.0 & 0.5 & 1.0 & 1.5 & 2.0 & 2.5 & 3.0  \\
\hline
 0 & 0.00,2.30 & 0.00,2.30 & 0.00,2.30 & 0.00,2.30 & 0.00,2.30  & 0.00,2.30 & 0.00,2.30   \\
 1 & 0.08,3.95 & 0.00,3.50 & 0.00,3.27 & 0.00,3.11 & 0.00,2.99  & 0.00,2.91 & 0.00,2.84  \\
 2 & 0.45,5.45 & 0.00,4.82 & 0.00,4.43 & 0.00,4.11 & 0.00,3.87  & 0.00,3.67 & 0.00,3.52  \\
 3 & 0.95,6.91 & 0.45,6.40 & 0.00,5.70 & 0.00,5.28 & 0.00,4.92  & 0.00,4.62 & 0.00,4.36 \\
 4 & 1.52,8.33 & 1.01,7.84 & 0.53,7.29 & 0.09,6.60 & 0.00,6.08  & 0.00,5.69 & 0.00,5.34  \\
 5 & 2.14,9.70 & 1.63,9.21 & 1.14,8.71 & 0.65,8.15 & 0.22,7.48  & 0.00,6.85 & 0.00,6.44  \\
 6 & 2.79,11.05& 2.29,10.55& 1.79,10.05& 1.29,9.54 &0.81,8.98   & 0.37,8.34 & 0.00,7.60 \\
 7 & 3.47,12.36& 2.97,11.86& 2.47,11.37& 1.97,10.86& 1.48,10.35 & 1.00,9.80 & 0.55,9.18  \\
 8 & 4.18,13.66& 3.67,13.16& 3.17,12.66& 2.67,12.16& 2.17,11.66 & 1.68,11.14& 1.20,10.60  \\
 9 & 4.90,14.93& 4.39,14.43& 3.90,13.93& 3.39,13.43& 2.89,12.94 & 2.40,12.43& 1.90,11.92 \\
10 & 5.63,16.19& 5.13,15.69& 4.63,15.20& 4.13,14.70& 3.63,14.20 & 3.13,13.70& 2.63,13.19 \\
11 & 6.38,17.44& 5.88,16.95& 5.38,16.44& 4.88,15.95& 4.38,15.45 & 3.88,14.95& 3.38,14.44 \\
12 & 7.14,18.68& 6.64,18.18& 6.14,17.68& 5.64,17.18& 5.14,16.69 & 4.64,16.18& 4.14,15.69 \\
13 & 7.91,19.91& 7.41,19.41& 6.91,18.91& 6.41,18.41& 5.91,17.92 & 5.41,17.41& 4.91,16.91 \\
14 & 8.69,21.13& 8.19,20.63& 7.69,20.13& 7.19,19.63& 6.69,19.14 & 6.19,18.63& 5.68,18.13 \\
15 & 9.47,22.35& 8.97,21.85& 8.47,21.35& 7.97,20.85& 7.47,20.35 & 6.97,19.85& 6.47,19.35 \\
16 &10.26,23.55& 9.77,23.05& 9.26,22.55& 8.77,22.05& 8.26,21.55 & 7.76,21.05& 7.27,20.56 \\
17 &11.06,24.75&10.57,24.25&10.06,23.75& 9.57,23.25& 9.06,22.75 & 8.56,22.25& 8.06,21.75  \\
18 &11.87,25.95&11.37,25.45&10.87,24.95&10.37,24.45& 9.87,23.95 & 9.37,23.45& 8.87,22.95  \\
19 &12.68,27.14&12.18,26.64&11.68,26.14&11.18,25.64&10.68,25.14 &10.18,24.64& 9.68,24.14  \\
20 &13.50,28.32&13.00,27.82&12.49,27.32&11.99,26.82&11.49,26.32 &10.99,25.82&10.49,25.33  \\
\hline
\hline
$n\backslash b$ & 3.5 & 4.0 & 5.0 & 6.0 & 7.0 & 8.0 & 9.0  \\
\hline
 0 & 0.00,2.30 & 0.00,2.30 & 0.00,2.30 & 0.00,2.30 & 0.00,2.30 & 0.00,2.30 & 0.00,2.30  \\
 1 & 0.00,2.78 & 0.00,2.74 & 0.00,2.67 & 0.00,2.62 & 0.00,2.58 & 0.00,2.55 & 0.00,2.53 \\
 2 & 0.00,3.39 & 0.00,3.29 & 0.00,3.12 & 0.00,3.00 & 0.00,2.92 & 0.00,2.85 & 0.00,2.79 \\
 3 & 0.00,4.15 & 0.00,3.97 & 0.00,3.68 & 0.00,3.48 & 0.00,3.32 & 0.00,3.20 & 0.00,3.10 \\
 4 & 0.00,5.04 & 0.00,4.78 & 0.00,4.36 & 0.00,4.04 & 0.00,3.80 & 0.00,3.61 & 0.00,3.46 \\
 5 & 0.00,6.06 & 0.00,5.72 & 0.00,5.15 & 0.00,4.71 & 0.00,4.37 & 0.00,4.10 & 0.00,3.89 \\
 6 & 0.00,7.16 & 0.00,6.76 & 0.00,6.06 & 0.00,5.49 & 0.00,5.04 & 0.00,4.67 & 0.00,4.38 \\
 7 & 0.15,8.46 & 0.00,7.88 & 0.00,7.07 & 0.00,6.37 & 0.00,5.80 & 0.00,5.34 & 0.00,4.96 \\
 8 & 0.75,9.99 & 0.33,9.31 & 0.00,8.15 & 0.00,7.35 & 0.00,6.67 & 0.00,6.09 & 0.00,5.62 \\
 9 & 1.42,11.38& 0.96,10.79& 0.15,9.41 & 0.00,8.40 & 0.00,7.62 & 0.00,6.95 & 0.00,6.37 \\
10 & 2.14,12.68& 1.66,12.14& 0.75,10.95& 0.00,9.51 & 0.00,8.65 & 0.00,7.88 & 0.00,7.21 \\
11 & 2.88,13.94& 2.39,13.43& 1.43,12.34& 0.57,11.08& 0.00,9.73 & 0.00,8.89 & 0.00,8.13 \\
12 & 3.64,15.18& 3.14,14.68& 2.16,13.64& 1.23,12.51& 0.40,11.19& 0.00,9.95 & 0.00,9.11 \\
13 & 4.41,16.42& 3.91,15.91& 2.92,14.90& 1.95,13.84& 1.05,12.66& 0.26,11.28& 0.00,10.16\\
14 & 5.19,17.64& 4.69,17.13& 3.69,16.13& 2.70,15.10& 1.75,14.02& 0.88,12.79& 0.12,11.36\\
15 & 5.97,18.85& 5.47,18.35& 4.47,17.35& 3.48,16.34& 2.50,15.30& 1.57,14.17& 0.73,12.90\\
16 & 6.76,20.05& 6.26,19.55& 5.26,18.55& 4.27,17.55& 3.27,16.53& 2.31,15.47& 1.40,14.31\\
17 & 7.56,21.26& 7.06,20.75& 6.06,19.75& 5.06,18.76& 4.07,17.75& 3.08,16.72& 2.13,15.64\\
18 & 8.37,22.45& 7.87,21.95& 6.87,20.95& 5.87,19.95& 4.87,18.95& 3.88,17.93& 2.90,16.89\\
19 & 9.18,23.64& 8.68,23.14& 7.68,22.14& 6.68,21.14& 5.68,20.14& 4.68,19.13& 3.69,18.11 \\
20 & 9.99,24.82& 9.49,24.32& 8.49,23.32& 7.49,22.32& 6.49,21.32& 5.50,20.32& 4.50,19.31\\
\hline
\hline
$n\backslash b$  & 10.0 & 11.0& 12.0 & 13.0 & 14.0 & 15.0 \\
\hline
 0 & 0.00,2.30
   & 0.00,2.30 & 0.00,2.30 & 0.00,2.30 & 0.00,2.30 & 0.00,2.30  \\
 1 & 0.00,2.51
   & 0.00,2.49 & 0.00,2.48 & 0.00,2.46 & 0.00,2.45 & 0.00,2.44  \\
 2 & 0.00,2.74
   & 0.00,2.71 & 0.00,2.67 & 0.00,2.65 & 0.00,2.62 & 0.00,2.60  \\
 3 & 0.00,3.02
   & 0.00,2.96 & 0.00,2.90 & 0.00,2.86 & 0.00,2.82 & 0.00,2.78  \\
 4 & 0.00,3.34
   & 0.00,3.24 & 0.00,3.16 & 0.00,3.09 & 0.00,3.03 & 0.00,2.98  \\
 5 & 0.00,3.71
   & 0.00,3.57 & 0.00,3.46 & 0.00,3.36 & 0.00,3.27 & 0.00,3.20  \\
 6 & 0.00,4.15
   & 0.00,3.95 & 0.00,3.80 & 0.00,3.66 & 0.00,3.55 & 0.00,3.45   \\
 7 & 0.00,4.65
   & 0.00,4.40 & 0.00,4.18 & 0.00,4.01 & 0.00,3.86 & 0.00,3.74   \\
 8 & 0.00,5.23
   & 0.00,4.90 & 0.00,4.63 & 0.00,4.41 & 0.00,4.22 & 0.00,4.06 \\
 9 & 0.00,5.89
   & 0.00,5.48 & 0.00,5.15 & 0.00,4.86 & 0.00,4.62 & 0.00,4.42 \\
10 & 0.00,6.63
   & 0.00,6.14 & 0.00,5.73 & 0.00,5.38 & 0.00,5.08 & 0.00,4.83 \\
11 & 0.00,7.46
   & 0.00,6.88 & 0.00,6.39 & 0.00,5.96 & 0.00,5.60 & 0.00,5.30  \\
12 & 0.00,8.36
   & 0.00,7.70 & 0.00,7.12 & 0.00,6.62 & 0.00,6.19 & 0.00,5.82  \\
13 & 0.00,9.33
   & 0.00,8.59 & 0.00,7.93 & 0.00,7.35 & 0.00,6.84 & 0.00,6.41 \\
14 & 0.00,10.36
   & 0.00,9.55 & 0.00,8.81 & 0.00,8.15 & 0.00,7.57 & 0.00,7.06 \\
15 & 0.00,11.43
   & 0.00,10.56& 0.00,9.75 & 0.00,9.02 & 0.00,8.36 & 0.00,7.78 \\
16 & 0.59,13.00
   & 0.00,11.61& 0.00,10.75& 0.00,9.95 & 0.00,9.22 & 0.00,8.57  \\
17 & 1.25,14.44
   & 0.46,13.09& 0.00,11.79& 0.00,10.94& 0.00,10.15& 0.00,9.42  \\
18 & 1.97,15.78
   & 1.10,14.56& 0.34,13.17& 0.00,11.97& 0.00,11.12& 0.00,10.34 \\
19 & 2.73,17.06
   & 1.81,15.92& 0.97,14.66& 0.23,13.25& 0.00,12.14& 0.00,11.30 \\
20 & 3.52,18.29
   & 2.56,17.21& 1.66,16.05& 0.84,14.76& 0.12,13.32& 0.00,12.31\\
\bottomrule
\end{tabular*}%
\end{center}

\clearpage

\vspace{3mm}

\begin{center}
\caption{Table~4.\quad 95\% $CL$.}
\vspace{-3mm}
\footnotesize
\begin{tabular*}{170mm}{c@{\extracolsep{\fill}}ccccccc}
\toprule
$n\backslash b$ & 0.0 & 0.5 & 1.0 & 1.5 & 2.0 & 2.5 & 3.0  \\
\hline
 0 & 0.00,3.00 & 0.00,3.00 & 0.00,3.00 & 0.00,3.00 & 0.00,3.00   & 0.00,3.00 & 0.00,3.00   \\
 1 & 0.04,4.75 & 0.00,4.35 & 0.00,4.11 & 0.00,3.94 & 0.00,3.81   & 0.00,3.72 & 0.00,3.64  \\
 2 & 0.31,6.33 & 0.00,5.77 & 0.00,5.38 & 0.00,5.06 & 0.00,4.80   & 0.00,4.60 & 0.00,4.43 \\
 3 & 0.73,7.87 & 0.23,7.36 & 0.00,6.75 & 0.00,6.34 & 0.00,5.97   & 0.00,5.66 & 0.00,5.39 \\
 4 & 1.22,9.38 & 0.72,8.89 & 0.24,8.31 & 0.00,7.67 & 0.00,7.23   & 0.00,6.83 & 0.00,6.48 \\
 5 & 1.77,10.83& 1.27,10.34& 0.77,9.83 & 0.30,9.22 & 0.00,8.53   & 0.00,8.08 & 0.00,7.66 \\
 6 & 2.36,12.23& 1.86,11.74& 1.36,11.25& 0.86,10.72& 0.40,10.12  & 0.00,9.36 & 0.00,8.90   \\
 7 & 2.98,13.61& 2.48,13.12& 1.98,12.62& 1.48,12.12& 0.99,11.59  & 0.53,10.99& 0.11,10.27 \\
 8 & 3.63,14.96& 3.13,14.47& 2.63,13.97& 2.13,13.47& 1.63,12.96  & 1.14,12.44& 0.67,11.85 \\
 9 & 4.30,16.29& 3.80,15.79& 3.30,15.30& 2.79,14.80& 2.29,14.30  & 1.80,13.79& 1.31,13.26 \\
10 & 4.98,17.60& 4.48,17.10& 3.98,16.61& 3.48,16.11& 2.98,15.61  & 2.48,15.11& 1.98,14.60\\
11 & 5.68,18.90& 5.18,18.40& 4.68,17.90& 4.18,17.40& 3.68,16.91  & 3.18,16.40& 2.68,15.90\\
12 & 6.40,20.18& 5.90,19.69& 5.40,19.19& 4.90,18.69& 4.40,18.19  & 3.90,17.69& 3.40,17.19 \\
13 & 7.12,21.46& 6.62,20.96& 6.12,20.46& 5.62,19.96& 5.12,19.46  & 4.62,18.96& 4.12,18.46 \\
14 & 7.86,22.72& 7.36,22.22& 6.86,21.72& 6.36,21.22& 5.86,20.72  & 5.36,20.22& 4.86,19.72 \\
15 & 8.60,23.97& 8.11,23.47& 7.61,22.98& 7.10,22.47& 6.61,21.98  & 6.11,21.48& 5.60,20.98 \\
16 & 9.36,25.22& 8.86,24.72& 8.36,24.22& 7.86,23.72& 7.36,23.22  & 6.86,22.72& 6.36,22.22 \\
17 &10.12,26.46& 9.62,25.96& 9.12,25.46& 8.62,24.96& 8.12,24.46  & 7.62,23.96& 7.12,23.46 \\
18 &10.89,27.69&10.39,27.19& 9.89,26.69& 9.39,26.19& 8.89,25.69  & 8.39,25.19& 7.89,24.69 \\
19 &11.66,28.92&11.16,28.41&10.66,27.91&10.16,27.41& 9.66,26.92  & 9.16,26.42& 8.66,25.92  \\
20 &12.44,30.14&11.94,29.64&11.44,29.14&10.94,28.64&10.44,28.14  & 9.94,27.64& 9.44,27.14\\
\hline
\hline
$n\backslash b$ & 3.5 & 4.0 & 5.0 & 6.0 & 7.0 & 8.0 & 9.0  \\
\hline
 0 & 0.00,3.00 & 0.00,3.00 & 0.00,3.00 & 0.00,3.00 & 0.00,3.00 & 0.00,3.00 & 0.00,3.00  \\
 1 & 0.00,3.58 & 0.00,3.53 & 0.00,3.45 & 0.00,3.39 & 0.00,3.34 & 0.00,3.31 & 0.00,3.28  \\
 2 & 0.00,4.29 & 0.00,4.17 & 0.00,3.99 & 0.00,3.86 & 0.00,3.75 & 0.00,3.67 & 0.00,3.60  \\
 3 & 0.00,5.16 & 0.00,4.97 & 0.00,4.66 & 0.00,4.42 & 0.00,4.24 & 0.00,4.10 & 0.00,3.99  \\
 4 & 0.00,6.16 & 0.00,5.89 & 0.00,5.43 & 0.00,5.09 & 0.00,4.82 & 0.00,4.60 & 0.00,4.43 \\
 5 & 0.00,7.27 & 0.00,6.92 & 0.00,6.33 & 0.00,5.85 & 0.00,5.48 & 0.00,5.18 & 0.00,4.94 \\
 6 & 0.00,8.46 & 0.00,8.05 & 0.00,7.33 & 0.00,6.73 & 0.00,6.24 & 0.00,5.84 & 0.00,5.52  \\
 7 & 0.00,9.70 & 0.00,9.25 & 0.00,8.42 & 0.00,7.70 & 0.00,7.10 & 0.00,6.60 & 0.00,6.18 \\
 8 & 0.25,11.16& 0.00,10.48& 0.00£¬9.57& 0.00,8.76 & 0.00,8.05 & 0.00,7.45 & 0.00,6.94 \\
 9 & 0.84,12.69& 0.41,12.03& 0.00,10.77& 0.00,9.89 & 0.00,9.09 & 0.00,8.38 & 0.00,7.77 \\
10 & 1.50,14.07& 1.02,13.51& 0.18,12.15& 0.00,11.05& 0.00,10.18& 0.00,9.39 & 0.00,8.69 \\
11 & 2.19,15.40& 1.70,14.87& 0.77,13.71& 0.00,12.25& 0.00,11.33& 0.00,10.47& 0.00,9.69 \\
12 & 2.90,16.69& 2.40,16.18& 1.43,15.11& 0.55,13.87& 0.00,12.50& 0.00,11.59& 0.00,10.74\\
13 & 3.62,17.96& 3.12,17.46& 2.14,16.44& 1.19,15.32& 0.35,13.99& 0.00,12.75& 0.00,11.85\\
14 & 4.36,19.22& 3.86,18.72& 2.86,17.72& 1.89,16.67& 0.97,15.50& 0.18,14.09& 0.00,12.99 \\
15 & 5.10,20.48& 4.60,19.98& 3.60,18.98& 2.61,17.96& 1.65,16.88& 0.77,15.65& 0.02,14.17\\
16 & 5.86,21.72& 5.36,21.22& 4.36,20.22& 3.36,19.22& 2.38,18.18& 1.44,17.07& 0.59,15.78\\
17 & 6.62,22.96& 6.12,22.46& 5.12,21.46 & 4.12,20.46& 3.13,19.44& 2.16,18.39& 1.24,17.23\\
18 & 7.39,24.19& 6.88,23.69& 5.89,22.69& 4.89,21.69& 3.89,20.69& 2.90,19.66& 1.94,18.59 \\
19 & 8.16,25.42& 7.66,24.92& 6.66,23.92& 5.66,22.92& 4.66,21.92& 3.66,20.91& 2.69,19.87 \\
20 & 8.94,26.63& 8.44,26.14& 7.44,25.14& 6.44,24.14& 5.44,23.14& 4.44,22.13& 3.45,21.12\\
\hline
\hline
$n\backslash b$  & 10.0 & 11.0& 12.0 & 13.0 & 14.0 & 15.0 \\
\hline
 0 & 0.00,3.00
   & 0.00,3.00 & 0.00,3.00 & 0.00,3.00 & 0.00,3.00 & 0.00,3.00  \\
 1 & 0.00,3.25
   & 0.00,3.23 & 0.00,3.22 & 0.00,3.20 & 0.00,3.19 & 0.00,3.18  \\
 2 & 0.00,3.55
   & 0.00,3.50 & 0.00,3.46 & 0.00,3.43 & 0.00,3.40 & 0.00,3.38  \\
 3 & 0.00,3.89
   & 0.00,3.81 & 0.00,3.75 & 0.00,3.69 & 0.00,3.64 & 0.00,3.60 \\
 4 & 0.00,4.29
   & 0.00,4.17 & 0.00,4.07 & 0.00,3.98 & 0.00,3.91 & 0.00,3.85 \\
 5 & 0.00,4.73
   & 0.00,4.57 & 0.00,4.43 & 0.00,4.31 & 0.00,4.22 & 0.00,4.13 \\
 6 & 0.00,5.25
   & 0.00,5.03 & 0.00,4.85 & 0.00,4.69 & 0.00,4.55 & 0.00,4.44  \\
 7 & 0.00,5.84
   & 0.00,5.55 & 0.00,5.31 & 0.00,5.11 & 0.00,4.94 & 0.00,4.78  \\
 8 & 0.00,6.51
   & 0.00,6.15 & 0.00,5.84 & 0.00,5.58 & 0.00,5.36 & 0.00,5.17 \\
 9 & 0.00,7.25
   & 0.00,6.81 & 0.00,6.44 & 0.00,6.12 & 0.00,5.84 & 0.00,5.61 \\
10 & 0.00,8.08
   & 0.00,7.56 & 0.00,7.11 & 0.00,6.72 & 0.00,6.38 & 0.00,6.10 \\
11 & 0.00,8.99
   & 0.00,8.38 & 0.00,7.85 & 0.00,7.39 & 0.00,6.99 & 0.00,6.64  \\
12 & 0.00,9.97
   & 0.00,9.27 & 0.00,8.66 & 0.00,8.12 & 0.00,7.65 & 0.00,7.25  \\
13 & 0.00,11.01
   & 0.00,10.24& 0.00,9.55 & 0.00,8.93 & 0.00,8.39 & 0.00,7.91  \\
14 & 0.00,12.10
   & 0.00,11.26& 0.00,10.50& 0.00,9.81 & 0.00,9.20 & 0.00,8.65  \\
15 & 0.00,13.22
   & 0.00,12.34& 0.00,11.51& 0.00,10.75& 0.00,10.07& 0.00,9.45  \\
16 & 0.00,14.37
   & 0.00,13.45& 0.00,12.57& 0.00,11.75& 0.00,11.00& 0.00,10.31  \\
17 & 0.43,15.89
   & 0.00,14.59& 0.00,13.67& 0.00,12.80& 0.00,11.99& 0.00,11.24  \\
18 & 1.05,17.38
   & 0.27,15.98& 0.00,14.79& 0.00,13.88& 0.00,13.02& 0.00,12.21  \\
19 & 1.75,18.76
   & 0.88,17.51& 0.14,16.07& 0.00,15.00& 0.00,14.10& 0.00,13.24  \\
20 & 2.48,20.06
   & 1.56,18.92& 0.72,17.63& 0.01,16.14& 0.00,15.20& 0.00,14.30  \\
\bottomrule
\end{tabular*}%
\end{center}

\begin{multicols}{2}

\begin{center}
\includegraphics{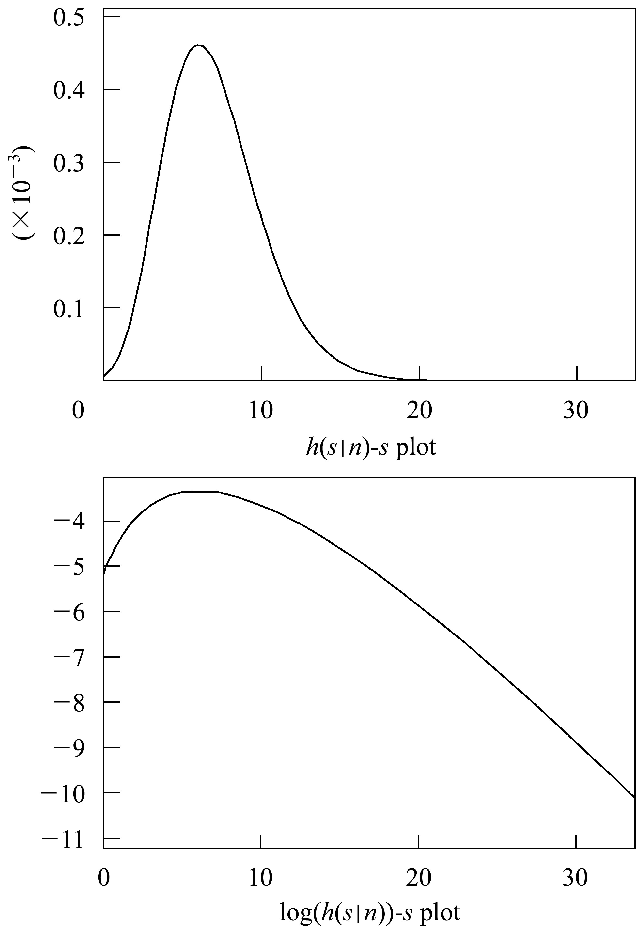}
\caption{Fig.~1.\quad The posterior density $h(s\vert n)$ as a function of $s$ for $n$=8, $ b$=2
with Gaussian systematic uncertainty $\sigma_b/b$=0.3 using uniform prior.}
\end{center}

\section{Summary}

The Bayesian HPD credible interval construction for a Poisson variable with
or without inclusion of the systematic uncertainties for background
expectation and signal efficiency has been illustrated. Such intervals have
several desirable properties. Like the unified interval, Bayesian HPD
intervals lie in the physical region and automatically change from credible
bounds to two-sided intervals. They avoid the drawback of background
dependence of unified intervals when zero events are observed. In addition,
the Bayesian HPD intervals are optimal on their own terms, and minimize the
length among all Bayesian intervals. Introducing the conditioning produces
shorter intervals in the case of incorporating the systematic uncertainty of
background expectation (and also the systematic uncertainty of signal
efficiency simultaneously), while keeps unchanged otherwise. Although the
Bayesian HPD intervals are conceptually different from the Frequentist
intervals, they are close numerically, that is, the frequentist coverage of
the Bayesian intervals is quite close to the Bayesian posterior credible
level\cite{9}. The code BPOCI provides a convenient tool to calculate the
Bayesian HPD intervals with or without conditioning and systematic
uncertainties.

\end{multicols}

\vspace{-2mm}
\centerline{\rule{80mm}{0.1pt}}
\vspace{2mm}

\begin{multicols}{2}


\end{multicols}

\clearpage


\begin{thebibliography}{90}
\bibitem{18} Consider that the $q(n\vert s)_b $ represented by Eqs.~(10), (11) is
$p(n\vert s)_b $ smeared by pdf of $f_{{b}'} (b,\sigma _b )$ and/or
$f_\varepsilon (1,\sigma _\varepsilon )$, in the case of $f$ being Gaussian,
Log-Gaussian or uniform (in limited region) pdf, the $h(s\vert n)$ with or
without inclusion of the systematic errors has similar shape as the function of $s$
\end{thebibliography}
\end{document}